\documentclass[aps,preprint,pra,groupedaddress,showpacs]{revtex4}
\usepackage{amssymb}
\usepackage{amsmath}
\usepackage{dcolumn}
\usepackage{bm}
\usepackage{graphicx}
\usepackage{mathrsfs}
\usepackage{appendix}

\begin{document}

\title{Preparing ground states and squeezed states of nanomechanical cantilevers by fast dissipation}
\author{Xin Wang, Hong-rong Li$^{\dag}$, Peng-bo Li, Chen-wei Jiang, Hong Gao, and Fu-li Li}
\affiliation{Department of Applied Physics, Xian Jiaotong University, Xian 710049, China}

\begin{abstract}
We propose a protocol that enables strong coupling between a flux qubit and the quantized motion of a magnetized nanomechanical cantilever. The flux qubit is driven by microwave fields with suitable parameters to induce sidebands, which will lead to the desired coupling. We show that the nanomechanical modes can be cooled to the ground states and the single-mode squeezed vacuum states can be generated via fast dissipation of the flux qubit. In our scheme, the qubit decay plays a positive role and can help drive the system to the target states.\newline
$^{\dag}$Corresponding Email: hrli@mail.xjtu.edu.cn\newline
\end{abstract}

\maketitle

\ \ \ \ \ \ \ \ \ \ \ \ \ \

%\keywords{}

\section{Introduction}

Quantum mechanics makes many breakthroughs that classical physics can not reach, but there exist huge obstacles when quantum theory is applied to the macroscopic systems. As one of the candidates for investigation of the macroscopic quantum effects, quantized motions of the mechanical oscillators of controllability have attracted much attention. Very earlier works on the quantized mechanical motions were proposed by considering the movable mirrors to show the quantum effects \cite{[1],[2],[3],[4],[5],[6]}. Owing to lack of effective ways to control the quantized mechanical motions, it is hard to show typical quantum phenomena in these systems. In recent years how to reach the quantum regime for the motion of massive objects has attracted great interests. For example, preparing the squeezed states of nanomechanical modes can be used to realized to the ultrahigh precision measurements \cite{[7]}. Since the amplitudes of the oscillation generated by the many sources of gravitational radiation are much smaller than the width of the ground statesÕ wave function, the detection of the gravity waves needs the massive detectors prepared in squeezed states to suppress the zero-point fluctuation of the ground states \cite{[8]}. Therefore, creating special quantum states (the quantum mechanics ground states, the squeezed states and the superposition states) for the mechanical modes \cite{[9],[10],[11],[12],[13]}, is one of the most challenging goals in macroscopic quantum systems. One way to conquer these difficulties is to find new protocols with the feature of controllability and enabling strong coupling of the quantized motions \cite{[14],[15],[16],[17]}. To enhance strong coupling for the mechanical modes, J. D. Teufel \emph{et al. }\cite{[18]} proposed a system in which a mechanical oscillator is be coupled to the circuit cavity, with the number of the driven photons are about 10$^{5}$, the coupling strength can reach \ 2$\pi \times 0.5$MHz. In another proposal demonstrated in \cite{[19]}, the authors showed a scheme that enables coupling strength about 115KHz between the cantilever and an electronic spin associated with a nitrogen-vacancy impurity in diamond.

In this work, we address the same problems by considering the coupling between a nanomechanical oscillator and a flux qubit. The protocol we proposed here is experimentally feasible, which enables strong coupling between the motion of a nanomechanical cantilever and a fast dissipative flux qubit. This hybrid system is coupled by the magnetic field gradient produced by a magnetic tip. There are two distinct features of our protocol. Firstly, the coupling strength can reach several MHz and is much stronger compared with some other protocols \cite{[9],[18],[19]}, and it can be used to realize quantum state transfer inside the hybrid systems. By employing flux driving to induce sidebands, and by adjusting different driving frequencies and coefficients we can obtain different coupling schemes (J-C coupling or anti-J-C coupling, driving on resonance or off-resonance) with controllable strength. Secondly, we can obtain the desired states of the mechanical motion through the fast dissipation of the qubit. As two examples based on our protocol, we will show how to prepare the nanomechanical mode to the ground states and the single-mode squeezed states by the dissipative processes. All these quantum states are important sources for testing the basic principles of quantum theory \cite{[20],[21],[22]}, and have many other applications such as the ultrahigh precision measurement in quantum metrology \cite{[7]}. Moreover, the idea shown in this work can be extended to a much larger system with a series of flux qubits and cantilevers.

This paper is organized as follows: In section II, the basic model of our scheme is presented. In section III, the cooling processes of the nanomechanical cantilevers based on our model is proposed and analyzed. In section IV, we show how to obtain squeezing for the nanomechanical cantilevers. The conclusion is made in section V.

\section{The Model of coupling a flux qubit to the quantized motions of a nanomechanical cantilever}

The idea of our work can be understood by considering a system shown in Fig. 1. We consider a hybrid quantum system that consists of a superconducting flux qubit coupled to a movable magnetic tip attached at the end of a cantilever, with the oscillating frequency $\omega $, and the $Q$ factor of the system is around 10$^{3}\sim $10$^{5}$ \cite{[23]}. For a cantilever with length $l$, width $w$ and thickness $t$, the fundamental frequency is calculated by $\omega =3.516\frac{t}{l^{2}}(\frac{E}{12\rho })^{2}$ under the condition $l\gg w\geq t$, where $E$ is the Young's modulus and $\rho$ is the density for the cantilever \cite{[23],[24]}. The motion of the tip will produce a time-varying magnetic field. The motion of the cantilever is described by the Hamiltonian $H_{M}=\hbar \omega b^{\dagger }b$, where $b^{\dagger }$($b$) are the creation (annihilation) operators for the quantized motion of the magnetic tip. Here we use a gap-tunable flux qubit with four Josephson junctions \cite{[25]}, for which the Hamiltonian is \cite{[26],[27]}
\begin{equation}
H_{Q}=\frac{1}{2}\epsilon (\Phi _{z})\bar{\sigma}_{z}+\Delta (\Phi _{x})\bar{%
\sigma}_{x},
\end{equation}
\begin{figure}[tbph]
\centering \includegraphics[width=10.0cm]{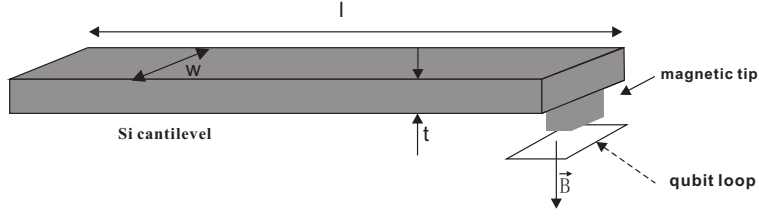}
\caption{The magnetic tip attached to the end of a nanomechanical cantilever with dimensions (l,w,t) is positioned above a flux qubit, thereby a strong coupling is achieved. The area of flux through the qubit can be considered as the same as the tip's size.}
\label{fig1}
\end{figure}
where $\bar{\sigma}_{z}=|+\rangle \langle +|-|-\rangle \langle -|$ and $\bar{\sigma}_{x}=|+\rangle \langle -|+|-\rangle \langle +|$ are the Pauli operators in the basis of clockwise $|+\rangle $ and anticlockwise $|-\rangle $ persistent currents. The energy bias satisfies $\epsilon (\Phi_{z})=2I_{p}(\Phi _{z}-\Phi _{0}/2)$, where $\Phi _{z}$ is the external flux threading the qubit loop, $I_{p}$ is the persistent current in the qubit loop, and $\Phi _{0}=\hbar /2e$ is the flux quantum.

The flux through the qubit loop can be expressed as $\Phi_{z}=\int_{S}(B_{total,z}(x,y,z)+B_{ext})dxdy\simeq \int_{S}[\frac{\partial
B_{total,z}(x,y,z)}{\partial z}|_{z=z_{0}}(z-z_{0})+B_{total,z}(x,y,z_{0})+B_{ext}]dxdy$, where $B_{total,z}(x,y,z)$ is the $z$-component of the total fields created by the
magnetic tip, while $B_{ext}$ is an applied external homogeneous magnetic filed, $S$ is the coupling area of the loop, $z_{0}$ is the average position of the
qubit loop, and $\frac{\partial B_{total,z}(x,y,z)}{\partial z}|_{z=z_{0}}$ is the magnetic field gradient at the position of the qubit. To reach the maximal magnetic coupling strength and minimize the flux noise, we suppose that the flux qubit is operated at the degeneracy point with the constant flux independent on the z position $\int_{S}(B_{total,z}(x,y,z_{0})+B_{ext})dxdy=\Phi _{0}/2$. Finally we obtain $\epsilon (\Phi _{z})=2I_{p}\int_{S}[\frac{\partial B_{total,z}(x,y,z)}{\partial z}|_{z=z_{0}}(z-z_{0})]dxdy$.

We use a type of magnetic tip with size $\sim 100$nm which was reported in \cite{[28]}. It is reasonable to consider that the magnetic gradient $\frac{\partial B_{total,z}(x,y,z)}{\partial z}|_{z=z_{0}}$ does not change very much in the area under the tip, that is $\frac{\partial B_{total,z}(x,y,z)}{\partial z}|_{z=z_{0}}\simeq G_{m}$, which can be considered as a constant in the effective coupling area $S_{eff}$. Therefore, the coupling term reads $\frac{1}{2}\epsilon (\Phi _{z})\bar{\sigma}_{z}=I_{p}G_{m}S_{eff}a_{0}(b^{\dagger }+b)\bar{\sigma}_{z}$, where $a_{0}(b^{\dagger }+b)=z-z_{0}$, and $a_{0}=\sqrt{\hbar /2m\omega }$ is the zero-point fluctuation amplitude of the mechanical oscillator with mass $m$. Now we can see that the first term of $H_{q}$ describes the coupling between the qubit and the nanomechanical cantilever, that is, $H_{int}=\hbar g_{0}\bar{\sigma}_{z}(b^{\dagger }+b)$, where the coupling strength $g_{0}=I_{p}S_{eff}G_{m}a_{0}/\hbar $. To estimate the coupling strength $g_{0}$, we adopt the following parameters \cite{[28],[29],[30],[31],[32]}: $I_{p}\approx $ 700nA, $a_{0}\approx $ 1.3$\times $10$^{-13}$m for a Si cantilever at a fundamental frequency $\omega\approx $ 2$\pi \times $50MHz, and $G_{m}\approx $ 8$\times $10$^{6}$T/m with $S_{eff}\approx $ 80nm$\times $80nm$\sim $0.64$\times $10$^{-14}$m$^{2}$. Then we find that $g_{0}$ is about 2$\pi \times $5.4MHz, which reaches the strong coupling regimes for the nanomechanical cantilevers. As can be seen from the expression for $g_{0}$, we can control the parameters of coupling area $S_{eff}$ and $G_{m}$, which are independent on the intrinsic parameters of the cantilever and the qubit. Another scheme for coupling a flux qubit to a nanomechanical oscillator is presented in \cite{[29]}, which can also reach the strong coupling regime via the Lorentz force. There a magnetic field about $\sim 5$mT is needed in order to create the coupling. Moreover, the nanomechanical oscillator is attached to the qubit loop rather than spatially independent from the qubit.

The qubit gap $\Delta (\Phi _{x})$ depends on the flux driving, which can be controlled independently by adjusting the magnetic flux $\Phi _{x}$ \cite{[28]}. In the present case, the qubit gap can be separated into a static part and a time-dependent part, i.e. $H_{d}(t)=\frac{1}{2}\hbar \nu\bar{\sigma}_{x}-\sum \eta _{i}\omega _{i}\hbar \cos (\omega _{i}t)\bar{\sigma}_{x}$, where $\eta _{i}$ are small coefficients and $\omega _{i}$ is the driving frequencies. In the new basis of the eigenstates of the qubit \{$|e\rangle =(|+\rangle +|-\rangle )/\sqrt{2}$, and $|g\rangle =(|+\rangle-|-\rangle )/\sqrt{2}$\}, the Hamiltonian of the hybrid system can be expressed as (setting $\hbar $ =1)
\begin{eqnarray}
H &=&H_{M}+H_{int}+H_{d}(t)  \notag \\
&=&\frac{1}{2}\nu \sigma _{z}+\omega b^{\dagger }b+g_{0}(\sigma _{+}+\sigma
_{-})(b^{\dagger }+b)-\sum \eta _{i}\omega _{i}\cos (\omega _{i}t)\sigma
_{z},
\end{eqnarray}%
where $\sigma _{+}=|e\rangle \langle g|$, $\sigma _{-}=|g\rangle \langle e|$, and in the new basis, the Pauli matrix $\sigma_z=|e\rangle \langle e|-|g\rangle \langle g|$.

\section{Cooling of a nanomechanical cantilever to the ground states by sideband transition}

Cooling of a mechanical resonator to its quantum ground states becomes a hot topic for various fields of physics, such as exploring the quantum mechanics on the macro-scales, ultrahigh precision measurements, and the detection of gravitational waves. In recent years, various theoretical works and experiments have devoted to cooling of the vibrational modes. For example, J. D. Teufel  \emph{et al.} \cite{[33]}, showed that a mechanical mode can also be parametrically coupled to a superconducting microwave resonant circuit, and at temperature of 15mK the mechanical mode of Q$\sim3.3\times 10^{5}$ can be cooled to a phonon occupation of 0.34$\pm $0.05.

In this section, we show in detail a cooling scheme for the ground states preparation of a cantilever based on the above setup. We consider a Si cantilever with dimensions ($l,w,t$)=(1.1,0.1,0.05)$\mu $m corresponding to a fundamental frequency $\omega \sim $2$\pi \times $50MHz. The working frequency of the flux qubit $\nu $ is in the range of tens of gigahertz. In order to cool the mechanical modes of the cantilever to its ground states, a flux driving $H_{dr}(t)=\eta \omega _{d}\cos (\omega _{d}t)\sigma _{z}$ is applied \cite{[34]}. The schematic diagram of the cooling process is described in Fig. 2.

\begin{figure}[tbph]
\centering \includegraphics[width=5.0cm]{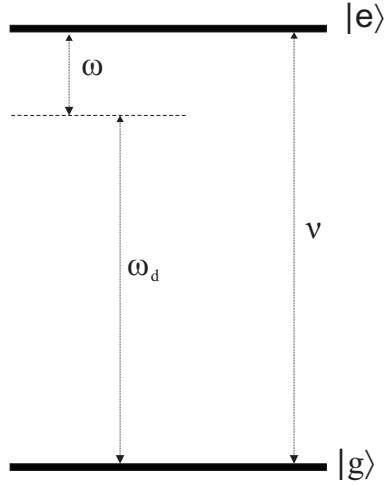}
\caption{Schematic of sideband cooling by the dissipation of a flux qubit. To select the J-C terms in Eq. (6), the flux qubit is driven red sideband with frequency $\protect\omega _{d}=\protect\nu -\protect\omega .$}
\label{fig2}
\end{figure}

According to Eq. (2), the total Hamiltonian for the whole system has the form

\begin{equation}
H_{1}=\frac{1}{2}\nu \sigma _{z}+\omega b^{\dagger }b+g_{0}(\sigma
_{+}+\sigma _{-})(b^{\dagger }+b)-\eta \omega _{d}\cos (\omega _{d}t)\sigma
_{z}.
\end{equation}%
Applying the unitary transformation $\hat{U}_{1}(t)=\exp (iH_{0}t)$ to $%
H_{1} $ with $H_{0}=\frac{1}{2}\nu \sigma _{z}+\omega b^{\dagger }b$, we
obtain
\begin{equation}
H_{2}=g_{0}(\sigma _{+}e^{i \nu t}+\sigma _{-}e^{-i \nu t})(b^{\dagger }e^{i\omega
t}+be^{-i\omega t})-\eta \omega _{d}\cos (\omega _{d}t)\sigma _{z}.
\end{equation}%
To go a further step, we proceed to perform another unitary transform,

\begin{eqnarray}
\hat{U}_{2}(t) &=&\hat{T}\exp [-i\int_{0}^{t}H_{dr}(t_{1})dt_{1}]  \notag \\
&=&1-i\int_{0}^{t}H_{dr}(t_{1})dt_{1}+(-i)^{2}\int_{0}^{t}dt_{1}%
\int_{0}^{t_{1}}H_{dr}(t_{1})H_{dr}(t_{2})dt_{2}+...
\end{eqnarray}%
where $\hat{T}$ denotes the time ordering operator. Keeping only the first order for small coefficient $\eta $ in the expanded form of $\hat{U}_{2}^{\dag }(t)\sigma _{+}\hat{U}_{2}(t)$ and $\hat{U}_{2}^{\dag}(t)\sigma _{-}\hat{U}_{2}(t)$, we obtain

\begin{equation}
H_{3}=g_{0}\sigma _{+}e^{i \nu t}[1-\eta (e^{i\omega _{d}t}-e^{-i\omega
_{d}t})](b^{\dagger }e^{i\omega t}+be^{-i\omega t})+\text{H.c.}
\end{equation}%
To achieve the cooling effect, as shown in Fig. 2, we need the driving is red sideband and select the J-C terms in $H_{3}$, i.e. $\nu -\omega_{d}=\omega $. If the condition \{$\nu ,\omega $\}$\gg g_{0}$ is satisfied, we can neglect rapidly oscillating terms of gigahertz and only keep terms of
megahertz. Then Eq. (6) reduces to the form
\begin{equation}
H_{4}=g(\sigma _{+}b+\sigma _{-}b^{\dagger })+g(\sigma _{+}b^{\dagger
}e^{2i\omega t}+\sigma _{-}be^{-2i\omega t}),
\end{equation}%
where $g=\eta g_{0}$. Hamiltonian $H_{4}$ allows the coherent state transfer between the flux quit and the cantilever. Note that the J-C term in the above equation represents the process contributing to cooling, while the anti-J-C term represents heating process.

Moreover, the mechanical modes can also decay to the thermal environment with finite temperature $T$, and the thermal phonon number at frequency $\omega $
can be express as $n_{th}=(e^{\hbar \omega /k_{B}T}-1)^{-1}\simeq k_{B}T/\hbar \omega$. Defining the Lindblad operator $D[A,\Omega ]\hat{\rho}%
=\frac{\Omega }{2}(A\hat{\rho}A^{+}-A^{+}A\hat{\rho})+$H.c. The master equation of density matrix $\hat{\rho}(t)$ for the qubit and the cantilever can be expressed as

\begin{equation}
\frac{d\hat{\rho}(t)}{dt}=-i[H_{4},\hat{\rho}(t)]+D[\sigma _{-},\Gamma ]\hat{%
\rho}(t)+n_{th}D[b^{\dag },\gamma ]\hat{\rho}(t)+(n_{th}+1)D[b,\gamma ]\hat{%
\rho}(t).
\end{equation}

In the equation above, $\gamma =$ $\omega /Q$ is the dissipative rate for the mechanical oscillator and $\Gamma $ represents the energy relaxation rate of the flux qubit. As demonstrated in appendix A, Under the condition $2g^{2}/\Gamma \ll \Gamma $, the excited states of the qubit can be adiabatically eliminated, thus we can obtain the evolution equation for the reduced density matrix $\mu $ of the cantilever as follows \cite{[35]}
\begin{equation}
\frac{d\mu }{dt}=D[b^{\dagger },\frac{2g^{2}}{\Gamma +2i\omega }%
+n_{th}\gamma ]\mu +D[b,\frac{2g^{2}}{\Gamma }+(n_{th}+1)\gamma ]\mu ,
\end{equation}%
where $A_{-}=2g^{2}/\Gamma $ represents the rate that the flux qubit brings out the phonons of the mechanical cantilever into the thermal environment, and real($A_{+})=2g^{2}\Gamma /(\Gamma ^{2}+4\omega ^{2})$ is the heating rate caused by the rotating-wave terms. In the equation above, $A_{+}+n_{th}\gamma $ is the total heating rate for the mechanical mode and $A_{-}+(n_{th}+1)\gamma $ is the cooling rate. If we assume that \{$\Gamma, g\}\ll \omega $, that is, real($A_{+})\ll n_{th}\gamma $ is satisfied, then $A_{+}$ can be neglected. let $\bar{n}(t)=$Tr$(\{\mu (t)b^{\dagger }b\}$ denote the mean phonon number of the mechanical cantilever, we can obtain a classical equation that describes the evolution for $\bar{n}$ \cite{[36]}:

\begin{equation}
\frac{d\bar{n}}{dt}=(\bar{n}+1)n_{th}\gamma -\bar{n}[\frac{2g^{2}}{\Gamma }
+(n_{th}+1)\gamma].
\end{equation}

It is easy to find from Eq. (10) that, the total effective cooling rate $\bar{n}[\frac{2g^{2}}{\Gamma }+(n_{th}+1)\gamma ]$ will decreases with $\bar{n},$ so there exits a cooling limit for this scheme. When the effective heating rate and cooling rate reach balance, the whole system reaches its steady states, which requires $\frac{d\bar{n}}{dt}=0$, thus we have:
\begin{equation}
\bar{n}=\frac{n_{th}\gamma }{\frac{2g^{2}}{\Gamma }+\gamma },
\end{equation}%
if $\frac{2g^{2}}{\Gamma }\gg \gamma $, the average phonon number is
\begin{equation}
\bar{n}=\frac{n_{th}\gamma \Gamma }{2g^{2}}\simeq \frac{k_{B}T\Gamma }{%
2\hbar Qg^{2}}.
\end{equation}%

Based on the quantum Monte Carlo simulations \cite{[37],[38]}, given that $\Gamma =2\pi \times 4$MHz \cite{[26],[27]}, $g=2\pi \times 1.0$MHz, $Q=5\times 10^{4}$, and $T=0.1K$, we obtain the numerical simulations of master equation (9) shown in Fig. 3. From the figure we find that, when the system reaches to the steady states, the average phonon number is about 0.06, which is in accordance with the theoretical results very well.

\begin{figure}[tbph]
\centering \includegraphics[width=10.0cm]{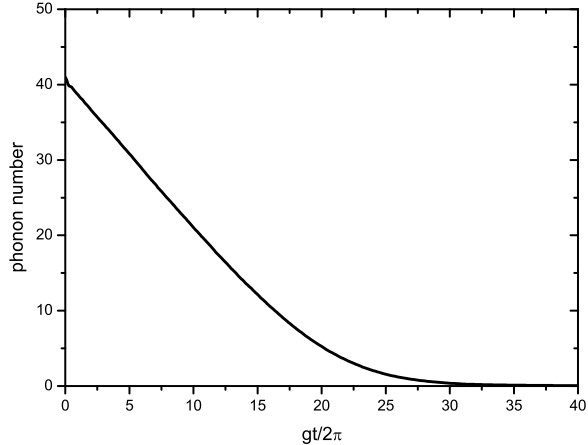}
\caption{The mean phonon number of the nanomechanical cantilever versus time
$gt/2\protect\pi $. Here T=0.1K, $n_{th}\simeq 41$, and $\protect\omega =50$
MHZ. }
\label{fig3}
\end{figure}

To understand the physical mechanism of the cooling process, we suppose that the temperature $T$ of the flux qubit is ranging from $1$K to $0.01$K initially. In such cryogenic temperature, the thermal boson number corresponding to the flux qubit $n_{q}=(e^{\hbar \nu /k_{B}T}-1)^{-1}\ll 1$, so the qubit tends to be in its ground states without the interaction with the motion of cantilever. However the thermal phonon number $n_{th}$ in the environment is quite large, and without the strong coupling of the qubit, the mechanical mode will evolve into the thermal states. Due to the strong cooling interaction between the flux qubit and the cantilever, phonons of the mechanical mode are taken away by the the flux qubit, which then dissipates quickly to the environment. As shown in Fig. 3, the mean phonon
number of the mechanical mode keeps reducing by the reiteration of the phonons transferring from the mechanical mode to the flux qubit until the cooling and heating rate reach the balance. From Eq. (12), we find that $\bar{n}$ decreases with the $Q$ factor and the coupling strength $g$, so we can enhance the cooling efficiency by increasing $Q$ and $g$.

\begin{figure}[tbph]
\centering \includegraphics[width=10.0cm]{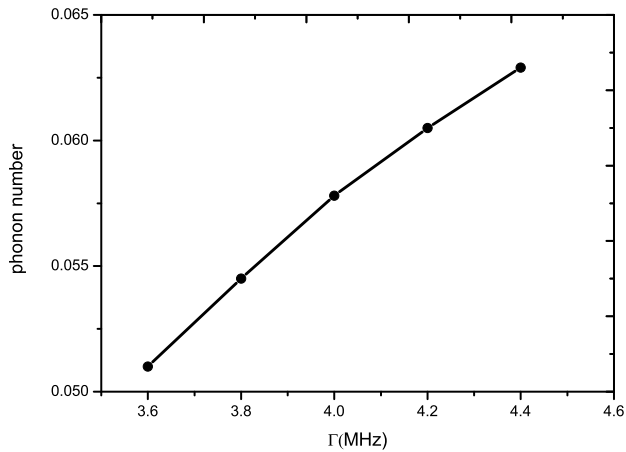}
\caption{The average phonon number changes with $\Gamma $ when the cooling
and heating rate reach the balance. The parameters are in the regime of $%
A^{+}\ll n_{th}\protect\gamma $ and $2g^{2}/\Gamma \ll \Gamma $.}
\label{fig4}
\end{figure}

In the regime $2g^{2}/\Gamma \ll \Gamma $, with very large dissipative rate of the flux qubit, the coherent transfer of phonons between the cantilever and the qubit will be destroyed by the qubit energy decay process with the environment. Thus the cooling rate $2g^{2}/\Gamma $ will decrease with $\Gamma $. As shown in Fig. 4, the average phonon number at steady state increases with the dissipative rate of the qubit. It means that very strong energy dissipative rate of the flux qubit will destroy the cooling process.

\section{Engineering the single-mode squeezed states of the mechanical cantilevers}

For a mechanical oscillator, even prepared in its ground state, there is still zero-point noise in both momentum and position. However, it is well known that, the fluctuation of position or momentum in the squeezed states can be smaller than that in the ground state, which is important in measurement requiring extremely high precision, for example, in detecting the effect of gravity wave. In this section, we propose a scheme to obtain the sinlge-mode squeezed states for the nanomechanical cantilevers.

The setup is shown in Fig. 1, where a nanomechanical cantilever with fundamental frequency $\omega $ is coupled to a flux qubit. We intend to achieve the single-mode squeezed states assisted with of dissipation of the qubit. In order to obtain higher mechanical frequency, we adopt a Si cantilever, for example, with dimensions (l,w,t)=(1,0.1,0.05)$\mu $m at frequency $\omega \simeq 2\pi \times 60$MHz.

The Hamiltonian for the system is
\begin{equation}
H_{s}=\frac{1}{2}\nu \sigma _{z}+\omega b^{\dagger }b+g(\sigma _{+}+\sigma
_{-})(b^{\dagger }+b)-[\eta ^{+}\omega ^{+}\cos (\omega ^{+}t)\sigma
_{z}+\eta ^{-}\omega ^{-}\cos (\omega ^{-}t)\sigma _{z}],
\end{equation}%
where $\eta ^{\pm }$ are the small coefficients for the driving frequencies $\omega ^{\pm }.$ Applying two unitary transformations $\hat{U}_{1}^{s}(t)=\exp (iH_{0}^{s}t)$ and $\hat{U}_{2}^{s}(t)=\hat{T}\exp[-i\int_{0}^{t}H_{2}^{s}(t)dt]$ with $H_{0}^{s}=\frac{1}{2}\nu \sigma_{z}+\omega b^{\dagger }b$ and $H_{2}^{s}(t)=-[\eta ^{+}\omega ^{+}\cos(\omega ^{+}t)\sigma _{z}+\eta ^{-}\omega ^{-}\cos (\omega ^{-}t)\sigma_{z}]$, respectively, and after performing the similar deduction in section III, we obtain a new form of the Hamiltonian
\begin{equation}
H_{q}=g\sigma _{+}e^{ivt}(b^{\dagger }e^{i\omega t}+be^{-i\omega t})\times
\{1-[\eta ^{+}(e^{i\omega ^{+}t}-e^{-i\omega ^{+}t})+\eta ^{-}(e^{i\omega
^{-}t}-e^{-i\omega ^{-}t})]\}+\text{H.c}.
\end{equation}%
In order to realize the squeezing processes, we suppose that the flux qubit is driven synchronously by a red sideband and a blue sideband flux drive. By setting $\omega ^{+}-v=v-\omega ^{-}=\omega $, we obtain two sideband couplings. The matching relations of the associated frequencies and energy level are shown in Fig. 5. If the condition \{$v,\omega ^{\pm},\omega $\}$\gg $\{$\eta ^{\pm }g$\} are satisfied, we can neglect the rapidly oscillating terms in Eq. (14) and obtain an effective form Hamiltonian
\begin{equation}
H_{eff}=\sigma _{+}(\eta ^{+}gb^{\dagger }+\eta ^{-}gb)+\text{H.c}.
\end{equation}

\begin{figure}[tbph]
\centering \includegraphics[width=5.0cm]{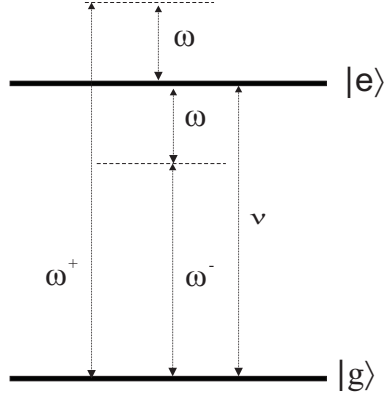}
\caption{Setup of preparing the single-mode squeezed states by the dissipation of the flux qubit. To select the desired coupling, the flux qubit is driven by two fluxes with different frequencies, $\protect\omega^{+}$ being the blue sideband driving frequency while $\protect\omega ^{-}$ being the red sideband driving frequency. }
\label{fig5}
\end{figure}
Let $\eta ^{-}g=\Theta _{1}$ and $\eta ^{+}g=\Theta _{2}$ in Eq. (15), and
the Hamiltonian becomes
\begin{equation}
H_{eff}=\sigma _{+}(\Theta _{1}b+\Theta _{2}b^{+})+\text{H.c}.
\end{equation}%
Based on Hamiltonian (16), the master equation of the density matrix $\bar{%
\rho}$ for the system is obtained:
\begin{equation}
\frac{d\bar{\rho}}{dt}=-i[H_{eff},\bar{\rho}]+D[\sigma _{-},\Gamma ]\bar{\rho}%
+n_{th}D[b^{\dag },\gamma ]\bar{\rho}+(n_{th}+1)D[b,\gamma ]\bar{\rho}.
\end{equation}%
By choosing $\zeta =\tanh ^{-1}(\Theta _{2}/\Theta _{1})$, $\Theta =\sqrt{\Theta _{1}^{2}-\Theta _{2}^{2}}$, $S(\zeta )=e^{\frac{1}{2}\zeta(b)^{2}-\frac{1}{2}\zeta (b^{+})^{2}}$, and applying a unitary transformation $\tilde{\rho}=S^{\dagger }(\zeta )\bar{\rho}S(\zeta )$, we obtain the following form of the master equation
\begin{equation}
\frac{d\tilde{\rho}}{dt}=-i[\tilde{H}_{eff},\tilde{\rho}]+D[\sigma _{-},\Gamma
]\tilde{\rho}+\Pi \lbrack \tilde{\rho}],
\end{equation}%
where $\tilde{H}_{eff}=\Theta (b\sigma _{+}+b^{+}\sigma _{-})$ and $\Pi\lbrack \tilde{\rho}]=n_{th}D[S^{\dagger }(\zeta )b^{\dag }S(\zeta ),\gamma ]\tilde{\rho}+(n_{th}+1)D[S^{\dagger }(\zeta )bS(\zeta ),\gamma ]\tilde{\rho}$.

As demonstrated in Appendix B, under the condtion that max\{$n_{th}\frac{\gamma }{2}+\frac{\gamma }{2}(2n_{th}+1)\sinh ^{2}\zeta ,(2n_{th}+1)\frac{\gamma }{2}\cosh \zeta \sinh \zeta \}\ll \Theta ,$ we can neglect terms in $\Pi \lbrack \tilde{\rho}].$ The master equation reduces to

\begin{equation}
\frac{d\tilde{\rho}}{dt}=-i[\tilde{H}_{eff},\tilde{\rho}]+D[\sigma _{-},\Gamma
]\tilde{\rho}.
\end{equation}
Apparently there is a steady state for the equation (19), that is
\begin{equation}
|\psi _{s}\rangle =|0\rangle |g\rangle .
\end{equation}
where $|g\rangle $ stands for the ground state of the \ flux qubit, and $|0\rangle $ represents the ground state of the mechanical mode. Reversing the unitary transformation, it can readily be seen that the steady state can be expressed as
\begin{equation}
|\Psi _{s}\rangle =S(\zeta )|0\rangle |g\rangle =e^{\frac{1}{2}\zeta (b)^{2}-%
\frac{1}{2}\zeta (b^{+})^{2}}|0\rangle |g\rangle,
\end{equation}
and $|\Psi _{s}\rangle $ is the unique stationary state of the master equation.

Obviously $e^{\frac{1}{2}\zeta (b)^{2}-\frac{1}{2}\zeta (b^{+})^{2}}|0\rangle $ is the single-mode squeezed states for the nanomechanical cantilever, and the degree of squeezing is determined by $\zeta =\tanh ^{-1}(\Theta _{2}/\Theta _{1})$. In order to achieve the maximally squeezed states, $\Theta _{2}\simeq \Theta _{1}$ must be satisfied, that corresponding to $\zeta \rightarrow \infty $. It also means that the effective coupling strength $\Theta \rightarrow 0$, i. e., it needs an extremely long time to reach the steady states. In this situation the squeezing process will be destroyed by the dissipation of the flux qubit and the mechanical modes caused by the thermal environment. So there is a balance between a large squeezing degree and the effective coupling strength.

\begin{figure}[tbph]
\centering \includegraphics[width=10.0cm]{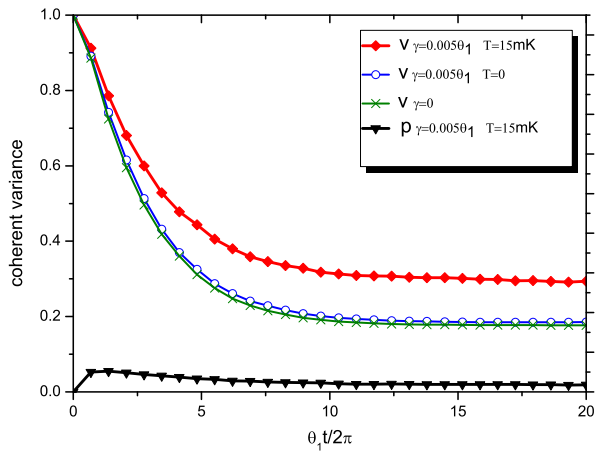}
\caption{(Color online) Coherent variance V of the the nanomechanical mode
and population P of the flux qubit in the excited states changes with
dimensionless variable $\Theta _{1}$t/2$\protect\pi $. We choose $\Gamma =2%
\protect\pi \times 4$MHz, $g=2\protect\pi \times 5$MHz, $\Theta _{1}=2%
\protect\pi \times 1$MHz, $\Theta _{2}=2\protect\pi \times 0.7$MHz
(corresponding $\protect\eta ^{-}=0.2$ and $\protect\eta ^{+}=0.14$), and $%
\protect\gamma $ =5KHz (corresponding to Q$=1.2\times 10^{4}$)}
\label{fig6}
\end{figure}
In fact, the dissipative rate for the flux qubit $\Gamma $ and the coupling strength $g$ can be several MHz, and the time for the system to reach the steady states is determined by both of $\Gamma $ and $\Theta $. It means that, when $\Gamma $ and $\Theta $ are comparable, the time to reach the steady states will be the order of a few times of max\{$\Theta ^{-1},\Gamma^{-1}\}$. Given $\Gamma \gg 2\Theta ^{2}/\Gamma $, we can adiabatically eliminate the excited states of the flux qubit similarily as shown in appendix A, and the rate to reach the steady states is $2\Theta^{2}/\Gamma $ \cite{[34],[39]}. To see more clearly, we show the numerical results based on the master equation (17) in the following, where we define the coherent variance for the nanomechanical mode as $V=\langle X^{2}\rangle-\langle X\rangle ^{2}$ with $X=b+b^{\dagger }$, and $V<1$ means squeezing \cite{[40]}. $P$ represents the population that the qubit is in its excited
states.

In Fig. 6, we plot evolution of the mechanical mode under different conditions. We find that the stationary squeezed states can be obtained when no thermal effects, i. e. $\gamma =0$. The squeezing is deteriorated when the temperature takes effects. However, the coherent variance is just slight destroyed when the temperature is up to about $\sim $10mK. Moreover, as we can see from the evolution of the population $P$, the flux qubit is always at very low excitations under the condition $\Gamma \gg 2\Theta ^{2}/\Gamma $. In Fig. 7, we show that the coherent variance will be destroyed with increasing of the temperature. At higher temperature with larger thermal phonon numbers, the squeezing is impacted by stronger dissipation by thermal environment. Therefore, our protocol might be promising in dilution
refrigerators of tens of millikelvin.
\begin{figure}[tbph]
\centering \includegraphics[width=10.0cm]{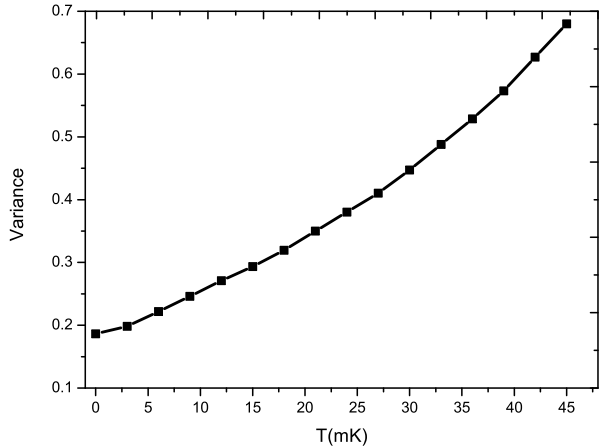}
\caption{The coherent variance as a function of the temperature. The parameters are the same with that in Fig. 6.}
\label{fig7}
\end{figure}

Finally, we note here two features of our scheme to generate squeezing of the nanomechanical cantilever: the first one is that, the steady squeezed states can always be produced because the squeezing process is assisted by the decay of the flux qubit. Secondly, the coupling between the qubit and the cantilever is controllable in a larger ranges, and the time to reach the steady squeezed states can be very short.

\section{Conclusion}

In conclusion, we have proposed an efficient scheme enabling strong coupling between a flux qubit and the quantized motion of a magnetized nanomechanical cantilever. In our scheme, the cantilever interacts with the flux qubit by the gradient magnetic field produced by a magnetic tip. With the flux qubit driven suitably, the motion of the cantilever can be controlled by the flux qubit. We have shown how to cool the cantilever to its ground states, and how to produce squeezed states of the motion of cantilever based on our scheme. The steady ground states and steady squeezed states can be obtained utilizing the fast dissipation of the flux qubit. Because the manufacture technology for the superconducting flux qubit and nanomechanical resonators is mature \cite{[29]}, our system provides possibilities to observe many quantum effects of the nanomechanical motions.

\section*{Acknowledgments}

This work was supported by the Natural Science Foundation of China under Grant Nos. 11174233 and 11104215.

\setcounter{equation}{0} \renewcommand{\theequation}{A\arabic{equation}}

\begin{appendix}

\section{Steps of adiabatically eliminating the excited states of the qubit}

We start from Eq. (7), i.e. the Hamiltonian between the flux qubit and the mechanical mode, we re-sign it as

\begin{equation}
H_{4}=g(\sigma _{+}e^{i\omega t}+\sigma _{-}e^{-i\omega t})(b^{\dagger
}e^{i\omega t}+be^{-i\omega t})
\end{equation}

We suppose that the density matrix for the system is $\hat{\rho}(t)=\mu (t)\otimes \rho _{q}(t)\otimes \rho _{B}$, where $\mu (t)$ represents the reduced density matrix for the mechanical modes, $\rho _{q}$ for the qubit, and $\rho _{B}$ for the thermal environment that the qubit decays to. Here we do not consider the terms for the mechanical modes interacting with the heat reservoir, which does not affect the deductions below. The interaction Hamiltonian between the qubit and the thermal environment is $V(t)=\sum_{k} \Gamma_{k}\sigma_{+}a_{k}e^{-i(v_{k}-v)t}+$H.c, where $a_{k}$ is the annihilation operator for the mode $k$ (corresponding with frequency $v_{k}$) of the thermal
environment. Thus the Hamiltonian for the whole system can be expressed as

\begin{equation}
H_{A1}=H_{4}+V(t).
\end{equation}

By assuming that the coupling strength $g$ between the qubit and the vibration mode is weak while the qubit interacts with the thermal environment at a strong rate $\Gamma$, then the qubit is dominated by the dissipative terms and the qubit is approximately in its ground state at all times, i.e. the density matrix is $\rho _{q}(t)\simeq |0\rangle \langle 0|$. Performing a transformation $\hat{U}(t)=\exp [-i\int_{0}^{t}V(t_{1})dt_{1}]$ to (A2), the Hamiltonian is then
\begin{equation}
H_{A2}=g\hat{U}^{\dag }(t)(\sigma _{+}e^{i\omega t}+\sigma _{-}e^{-i\omega t})%
\hat{U}(t)(b^{\dagger }e^{i\omega t}+be^{-i\omega t}),
\end{equation}%
and the master equation for the reduced density matrix $\mu (t)$ can be expressed as
\begin{equation}
\frac{d\mu (t)}{dt}=-\int_{0}^{t}dt^{^{\prime }}\text{tr}_{q}\text{tr}%
_{B}[H_{A2}(t),[H_{A2}(t^{^{\prime }}),\mu (t^{^{\prime }})\otimes \rho
_{q}(t^{^{\prime }})\otimes \rho _{B}],
\end{equation}%
where we have assumed that
\begin{equation}
\text{tr}_{q}\text{tr}_{B}[H_{A2}(t),\hat{\rho}(0)]=0,
\end{equation}%
and tr$_{q}$ and tr$_{B}$ means tracing over the qubit and the thermal environment. The Markov approximation is made by assuming that $\mu(t^{^{\prime }})$ does not change greatly during times 0$\rightarrow $t, which means the evolution time $\tau _{\mu }$ for $\mu
(t)$ being estimated as $\tau _{\mu }\gg t$, so we can replace $\mu (t^{^{\prime }})$ with $\mu (t)$,
\begin{equation}
\begin{split}
\frac{d\mu (t)}{dt}& =-\int_{0}^{t}dt^{^{\prime }}\text{tr}_{q}\text{tr}%
_{B}[H_{A2}(t),[H_{A2}(t^{^{\prime }}),\mu (t)\otimes \rho _{q}(t^{^{\prime }})\otimes
\rho _{B}] \\
& =\int_{0}^{t}dt^{^{\prime }}\text{tr}_{q}\text{tr}_{B}[H_{A2}(t^{^{\prime}})\mu (t)\otimes \rho _{q}\otimes \rho _{B}H_{A2}(t)-H_{A2}(t)H_{A2}(t^{^{\prime }})\mu(t)\otimes \rho _{q}(t^{^{\prime }})\otimes \rho _{B}]+\text{h.c.}.
\end{split}%
\end{equation}%
We re-express $H_{A2}$ as $H_{A2}=B(t)[b^{\dagger }e^{i\omega t}+be^{-i\omega t}]$, where $B(t)=$ $g[\sigma _{+}(t)e^{i\omega t}+\sigma _{-}(t)e^{-i\omega t}]$, and $\sigma _{+}(t)=\hat{U}^{\dag }(t)\sigma _{+}\hat{U}(t),\sigma _{-}(t)=\hat{U}^{\dag }(t)\sigma _{-}\hat{U}(t)$. Neglecting the oscillating terms in
Eq. (A6) and defining $t^{^{\prime }}=t-s$, we get
\begin{equation}
\begin{split}
\frac{d\mu (t)}{dt}& =\int_{0}^{t}ds\text{tr}_{q}\text{tr}_{B}[B(t-s)\rho
_{q}(t-s)\otimes \rho _{B}B(t)]e^{-i\omega s}b^{\dagger }\mu (t)b \\
& +\int_{0}^{t}ds\text{tr}_{q}\text{tr}_{B}[B(t-s)\rho _{q}(t-s)\otimes \rho
_{B}B(t)]e^{i\omega s}b\mu (t)b^{\dagger } \\
& -\int_{0}^{t}ds\text{tr}_{q}\text{tr}_{B}[B(t)B(t-s)\rho _{q}(t-s)\otimes
\rho _{B}]e^{i\omega s}b^{\dagger }b\mu (t) \\
& -\int_{0}^{t}ds\text{tr}_{q}\text{tr}_{B}[B(t)B(t-s)\rho _{q}(t-s)\otimes
\rho _{B}]e^{-i\omega s}bb^{\dagger }\mu (t)+\text{h.c.}.
\end{split}%
\end{equation}%
Let $\int_{0}^{t}ds$tr$_{q}$tr$_{B}[B(t-s)\rho _{q}(t-s)\otimes \rho_{B}B(t)]e^{-i\omega s}=\int_{0}^{t}ds$tr$_{q}$tr$_{B}[B(t)B(t-s)\rho
_{q}(t-s)\otimes \rho _{B}]e^{-i\omega s}=A_{+}(t)$, $\int_{0}^{t}dstr_{q}tr_{B}[B(t)B(t-s)\rho _{q}(t-s)\otimes \rho_{B}]e^{i\omega s}=\int_{0}^{t}dstr_{q}tr_{B}[B(t-s)\rho _{q}(t-s)\otimes\rho _{B}B(t)]e^{i\omega s}=A_{-}(t)$, we rewrite (A7) as
\begin{equation}
\frac{d\mu (t)}{dt}=A_{+}(t)[b^{\dagger }\mu (t)b+-bb^{\dagger }\mu
(t)]+A_{-}(t)[b\mu (t)b^{\dagger }-b^{\dagger }b\mu (t)]+\text{h.c.}.
\end{equation}%

Now we use some new signs as $\rho _{q,ee}(t)=$tr$_{q}[|e\rangle \langle e|\rho _{q}(t)],$ $\rho _{q,gg}(t)=$tr$_{q}[|g\rangle \langle g|\rho _{q}(t)]$ and $\rho _{q,ge}(t)=$tr$_{q}[|g\rangle \langle e|\rho _{q}(t)]$. Since the qubit dissipates very fast, dynamics of $\rho _{q,ge}(t)$ and $\rho _{q,eg}(t)$ can be approximately expressed as $\sigma _{-}(t)=\rho _{q,ge}(t)\simeq \rho_{q,ge}(t-s)e^{-\Gamma s},$ and $\sigma _{+}(t)=\rho _{q,eg}(t)\simeq \rho_{q,eg}(t-s)e^{-\Gamma s}$. Because of fast dissipation, the qubit is approximately in the ground states all the time, which means $\rho _{q,ee}(t-s)\simeq 0$ and $\rho _{q,gg}(t-s)\simeq 1$. Under the condition that t$\gg \Gamma ^{-1}$, and according the quantum regression theorem \cite{[41]}, $A_{+}(t)$ and $A_{-}(t)$ will reduce to the simple forms,
\begin{align}
A_{+}& =g^{2}\int_{0}^{t}\rho _{q,ee}(t-s)e^{-\Gamma
s}ds+g^{2}\int_{0}^{t}\rho _{q,gg}(t-s)e^{-\Gamma s}e^{-i2\omega s}ds\simeq
\frac{g^{2}}{\Gamma +2i\omega }, \\
A_{-}& =g^{2}\int_{0}^{t}\rho _{q,ee}(t-s)e^{-\Gamma s}e^{i2\omega
s}ds+g^{2}\int_{0}^{t}\rho _{q,gg}(t^{^{\prime }})e^{-\Gamma s}ds\simeq
\frac{g^{2}}{\Gamma },
\end{align}%
and the equation of $\mu (t)$ has the form

\begin{equation}
\begin{split}
\frac{d\mu (t)}{dt}& =\frac{g^{2}}{\Gamma }[2b\mu (t)b^{\dagger }-b^{\dagger
}b\mu (t)-\mu (t)b^{\dagger }b] \\
& +\frac{g^{2}}{\Gamma +2i\omega }[b^{\dagger }\mu (t)b-bb^{\dagger }\mu
(t)]+\frac{g^{2}}{\Gamma -2i\omega }[b^{\dagger }\mu (t)b-\mu (t)bb^{\dagger}].
\end{split}%
\end{equation}%

Eq. (A11) tell us that $\tau _{\mu }\sim $ $2\Gamma /g^{2}$, since the equation was obtained under the condition $\tau _{\mu }\gg t\gg \Gamma^{-1}$, we find that the condition for adiabatically eliminating the qubit's excited states is $\Gamma \gg \frac{2g^{2}}{\Gamma }$.

\setcounter{equation}{0} \renewcommand{\theequation}{B\arabic{equation}}

\section{Dissipative term for the mechanical mode after the squeezed transformation}

We start from Eq. (17). After applying a unitary transformation $\tilde{\rho}=S^{\dagger }(\zeta )\bar{\rho}S(\zeta )$, the dissipative terms for
the cantilever can be written as
\begin{equation}
\Pi \lbrack \tilde{\rho}]=n_{th}D[S^{\dagger }(\zeta )b^{\dag }S(\zeta
),\gamma ]\tilde{\rho}+(n_{th}+1)D[S^{\dagger }(\zeta )bS(\zeta ),\gamma ]%
\tilde{\rho}.
\end{equation}%
Using the relation $S^{\dagger }(\zeta )b^{\dag }S(\zeta )=b^{\dag }\cosh
\zeta -b\sinh \zeta $ ,$S^{\dagger }(\zeta )bS(\zeta )=b\cosh \zeta -b^{\dag
}\sinh \zeta $ and $\cosh ^{2}\zeta -\sinh ^{2}\zeta =1,$ we can obtain

\begin{equation}
\begin{split}
\Pi \lbrack \tilde{\rho}]& =n_{th}\frac{\gamma }{2}\{\cosh ^{2}\zeta
D[b^{\dag },\gamma ]\tilde{\rho}+\sinh ^{2}\zeta D[b,\gamma ]\tilde{\rho}%
-\cosh \zeta \sinh \zeta \Pi s[\tilde{\rho}]\}+ \\
& (n_{th}+1)\frac{\gamma }{2}\{\sinh ^{2}\zeta D[b^{\dag },\gamma ]\tilde{%
\rho}+\cosh ^{2}\zeta D[b,\gamma ]\tilde{\rho}-\cosh \zeta \sinh \zeta \Pi s[%
\tilde{\rho}]\} \\
& =[n_{th}\frac{\gamma }{2}+\frac{\gamma }{2}(2n_{th}+1)\sinh ^{2}\zeta
]D[b^{\dag },\gamma ]\tilde{\rho} \\
& +[(n_{th}+1)\frac{\gamma }{2}+\frac{\gamma }{2}(2n_{th}+1)\sinh ^{2}\zeta
]D[b,\gamma ]\tilde{\rho} \\
& -(2n_{th}+1)\frac{\gamma }{2}\cosh \zeta \sinh \zeta \Pi s[\tilde{\rho}],
\end{split}%
\end{equation}%
where $\Pi s[\tilde{\rho}]=(2b^{\dag }\tilde{\rho}b^{\dag }-b^{\dag }b^{\dag
}\tilde{\rho}-\tilde{\rho}b^{\dag }b^{\dag })+(2b\tilde{\rho}b-bb\tilde{\rho}%
-\tilde{\rho}bb).$

\end{appendix}


\begin{thebibliography}{99}
\bibitem{[38]} In the Monte Carlo simulations, we define the operators $%
C_{1}=\sqrt{\Gamma }\sigma _{-}$, $C_{2}=\sqrt{n_{th}\gamma }b^{\dag }$, $%
C_{3}=\sqrt{(n_{th}+1)\gamma }b$ to describe the quantum jumps. As
demonstrated in Ref. [33], the non-Hermitian Hamiltonian $H=H_{s}-\frac{i}{2}%
\sum_{m=1}^{3}C_{m}^{\dag }C_{m}$ \ and \{$C_{1}$,$C_{2}$,$C_{3}$\}
determine the whole process of the numerical simulation. In this artical,
the numbers of the simulation trajectories are 3000.
\end{thebibliography}
\end{document}